\documentclass[10pt,letterpaper,twocolumn]{article} 

\usepackage{ol2}
\usepackage[draft]{hyperref}
\usepackage{amsmath}
\usepackage{epsfig}

\begin{document}

\twocolumn[ 

\title{Discrete Cylindrical Vector Beam Generation from an Array of Optical Fibers}


\author{R. Steven Kurti$^{1,*}$, Klaus Halterman$^{1}$, Ramesh K. Shori,$^1$ and Michael J. Wardlaw$^2$}

\address{
$^1$Research  Division, Physics and Computational Sciences Branch, Naval Air Warfare Center,\\ China Lake, California, 93555, USA
\\
$^2$Office of Naval Research, Arlington, Virgina, 22203, USA. \\ 
$^*$Corresponding author: steven.kurti@navy.mil
}

\begin{abstract} 
A novel method is presented for the beam shaping of
far field intensity distributions of coherently 
combined fiber arrays. The fibers are arranged uniformly 
on the perimeter of a circle, and the linearly polarized beams
of equal shape are superimposed such that 
the far field 
pattern represents an effective radially 
polarized vector beam, or discrete cylindrical vector (DCV) beam.
The DCV beam is produced by 
three or more beams 
that each individually
have a varying polarization vector. 
The beams are appropriately distributed 
in the near field
such that the far field intensity distribution 
has a central null. 
This result is in contrast to the situation of 
parallel linearly polarized beams, 
where the intensity peaks on axis.
\end{abstract}


 ] 
\bigskip
\noindent
The propagation of electromagnetic fields
from multiple fibers can result in complex far-field intensity 
profiles that depend crucially on the individual near field  
polarization and phase. 
Coherently phased arrays have been used in defense and communications systems for many years,
where the antenna ensemble forms a diffraction pattern that can be altered by changing the antenna spacing, 
amplitude, and relative phase relationships. 
In the optical regime, similar systems have recently been formed by actively or passively locking the 
phases of two or more identical optical beams \cite{shay0,brus,simp,shir}.  
While coherently combined fiber lasers are increasingly gaining acceptance as sources for high power applications,
most approaches involve phasing a rectangular or hexagonal grid of fibers with collimated beams that are 
linearly polarized along the same axis \cite{shay}.  
The resultant far field diffraction pattern therefore is peaked 
in the center.  
If the polarization is allowed to vary, as it does in 
radial vector beams, the
diffraction pattern vanishes in the center\cite{oron}, which
can result in a myriad of device applications, discussed below.

Several types of polarization states have been investigated,
including radial and azimuthal \cite{mach,moser,oron,pass,tid,nest,hira,bomzon,yone,gros,wynne,kozawa,salamin}.  
These beams are used in  mitigating thermal effects in high power lasers \cite{moshe,moshe2,roth}, 
laser machining \cite{nizi,rioux}, and particle 
acceleration interactions \cite{salamin2,ift}.  
They can even be used to generate longitudinal 
electric fields when tightly focused \cite{young,dorn}, 
and although they are typically formed in free space 
laser cavities using a conical lens, or axicon, they can also be formed and guided in fibers \cite{gros2,volpe,li}.
Typical methods for creating radially polarized beams fall into two categories: the first involves 
an intracavity axicon in a laser resonator to generate the laser mode, the second begins with 
a single beam and rotates the polarization of portions of the beam to create an inhomogeneously 
(typically radially or azimuthally) polarized beam.  
In this Letter we 
demonstrate
a novel approach that takes an array of Gaussian beams,
each with the appropriately oriented linear polarization, 
and then superimpose them in the far-field, effectively creating 
a DCV beam.

Our starting point in determining the propagation of polarized
electromagnetic beams 
is
the vector Helmholtz equation, $\nabla \times \nabla\times {\bf E}({\boldsymbol \rho},z) - k_0^2 {\bf E}({\boldsymbol \rho},z) =0$.
We use cylindrical coordinates, so that 
$\boldsymbol \rho$ 
is the usual transverse coordinate, $k_0=\omega/c$, 
and the usual sinusoidal time dependence has
been factored out. 
Applying Lagrange's formula permits the wave equation  to be written in terms of the
vector Laplacian: $\nabla^2 {\bf E}({\boldsymbol \rho},z)+k_0^2 {\bf E}({\boldsymbol \rho},z) =0$,
which can then be reduced to the  paraxial wave equation.
The paraxial solutions are then inserted into the
familiar
Fraunhofer diffraction integral, which for propagation 
at sufficiently large $z$, 
yields the electric field,
\begin{align}
\label{fraun}
&{\bf E}(k_x,k_y) =  \exp\Bigl(\frac{i k_0 \rho^2}{2z}\Bigr) \frac{ \exp(i k_0 z)}{i \lambda z}  \nonumber \\
& \times \sum_{j=1}^{N}  \int_{0}^{a} r dr \int_{0}^{2\pi}  d\phi\,  e^{-i{\bf k}\cdot({\bf r}+{\bf s}_j)} 
{\bf E}^0({\bf r}+{\bf s}_j,0), 
\end{align}
where 
${\bf E}^0({\bf r}+{\bf s}_j,0)$ is the incident electric field in the 
plane of the fiber array. 
\begin{figure}[htb]\centerline{
\includegraphics[width=.25\textwidth]{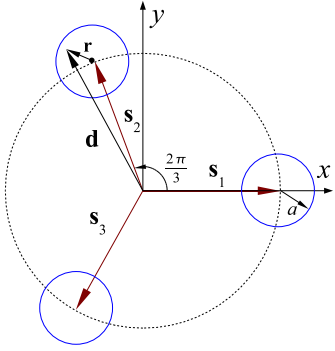}}
\caption{Example of the multiple fiber setup. A specific case of $N=3$ is shown, where
each beam is represented by a circle of radius $a$ and arranged
uniformly on a circle of radius $R$. Points within the $i$th hole are located 
by ${\bf d} = {\bf s}_i + {\bf r}$, with the vector ${\bf r}$ 
originating at the center of 
each hole.}
\label{fig0}
\end{figure}
It is clear that longitudinal polarization is not considered here, consistent with the paraxial approximation.
The far field intensity will be a complex diffraction pattern that depends on the individual beam intensity 
profile in the near field as well its polarization state and optical phase.  
A diagram illustrating the geometry used is shown in Fig.~\ref{fig0}, where
an array of three 
circular holes of radius $a$  are equally distributed on the circumference of a circle
of radius $R$.
The center of each hole is located at ${\bf s}_j=R \hat{\bf r}_j$,
where $\hat{\bf r}_j\equiv (\hat{\bf x}\cos\theta_j+\hat{\bf y}\sin\theta_j)$ is the unit radial vector,
and ${\bf r}$ is the coordinate relative to each center. The center of each fiber is separated by an angle $\theta_j=2\pi(j-1)/N$.
One can scale to any number of beams, constrained only by the radius $R$.
For identical Gaussian beams, 
the incident field is expressed as, 
${\bf E}^0 
={E}^0({\bf r})\hat{\bf r}_j$, 
where ${E}^0({\bf r})=E_0 e^{-r^2/w_0^2}$.
This permits
Eq.~(\ref{fraun}) to be 
separated,
\begin{align}
{\bf E}(k_x,k_y) = {\cal F}(\rho,z)\sum_{j=1}^{N}  e^{-i{\bf k}\cdot{\bf s}_j}  \hat{\bf r}_j, \label{int} 
\end{align} 
where  
${\cal F}(\rho,z) =E_0 k_0/z \int^{a}_{0} r dr \,J_0(k_0 r \rho/z)  e^{-r^2/w_0^2}$,
and the prefactors that do not contribute to the intensity have been suppressed.
It is clear from Eq.~(\ref{int}) that the far field transform preserves
the radially symmetric polarization state of the system, as expected.
Note that in the limit $w_0/a \gg 1$, 
and using the relation $x J_0(x)=[x J_1(x)]'$, we have ${\cal F}(\rho,z) \approx E_0 (a/\rho) J_1(k_0 a \rho/z) $,
which gives the expected Fraunhofer diffraction pattern for a circular aperture of radius $a$.
In the opposite limit, where the Gaussian profile is narrow enough so that the aperture geometry has little effect,
we have ${\cal F}(\rho,z) \approx E_0 k_0 w_0^2/(2z) \exp[-(k_0 \rho w_0/(2 z))^2]$.
For the multiple beam arrangements investigated, corresponding
to $N=3,4,6$, we then can calculate the intensity, $I_N$, explicitly:
\begin{align}
{\cal K}_3 (\phi) & =  3-\cos ({\sqrt{3}} k_y R)-2\cos (\frac{\sqrt{3}}{2} k_y R) \cos\Bigl(\frac{3}{2} k_x R \Bigr), \\
{\cal K}_4 (\phi)& = 4\bigl[\sin^2(k_x R)+\sin^2(k_y R)\bigr], \\
{\cal K}_6 (\phi)& = 4\Bigl[ \Bigl(\cos\Bigl(\frac{\sqrt{3}}{2} k_y R\Bigr)  \sin\Bigl(\frac{1}{2}k_x R\Bigr) + \sin\bigl(k_x R\bigr) \Bigr)^2 \nonumber \\ 
 & \,\,\,\,\,\, +3 \cos^2\Bigl(\frac{1}{2}k_x R\Bigr)\sin^2\Bigl(\frac{\sqrt{3}}{2}k_y R\Bigr)\Bigr], 
\end{align}
where 
$I_N\equiv {\cal F}^2(\rho,z) {\cal K}_N(\phi;\rho,z)$, 
$k_x = (k_0/z) \rho \cos\phi$ and $k_y = (k_0/z) \rho \sin\phi$. To
rotate the array, 
one can perform a standard rotation
${\bf r}' = {\cal R}(\phi'){\bf r}$, so e.g., a $\pi/4$ rotation would give, 
${\cal K}_4(\phi')= 4(1-\cos(\sqrt{2} k_x R)\cos(\sqrt{2} k_y R))$.

We employ two different methods 
to create the DCV beams previously described.
In each method, the experimental configuration
utilizes collimated Gaussian beams that are in phase.
The arrays are cylindrically symmetric both with respect
to intensity and polarization.
We focus the beams with a transform lens in order to simulate the
far field intensity at the lens focus. 
The focal spot is then imaged onto a camera by
a microscope objective in order to fill the CCD array.
The laser source is a $100$ mW continuous wave (CW) Nd:YAG 
operating at  $\lambda = 1.064\mu$m.

\begin{figure}[htb]\centerline{
\includegraphics[width=.5\textwidth]{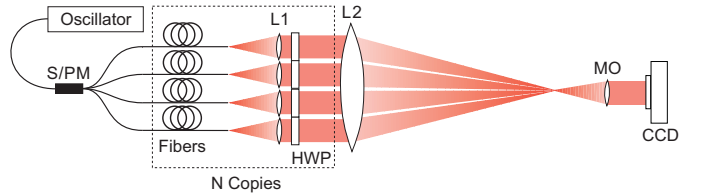}}
\caption{Diagram of the
experimental setup for DCV beam generation. The oscillator output is split and phase modulated (S/PM).  
The output of each fiber is collimated by means of a lens, L1, and then the polarization is 
rotated by a half-wave plate (HWP).  The beam ensemble is then focused to a point which is imaged 
by a microscope objective (MO) onto a charge-coupled-device (CCD) array.
}
\label{figexp}
\end{figure}
The first method (for the case of $N=4$) involves expanding
a beam from the Nd:YAG by means of
a telescope beam expander to a diameter of about 1cm. 
The 
diffractive optical element then
creates an $8\times8$ array of beams. Four of the beams are reflected and made nearly parallel by a segmented mirror, and then
linearly polarized upon passing through a set of four half-wave 
plates. The phases of each beam are made identical by having them pass through articulated glass slides.
\begin{figure}
\centering
\begin{tabular}{cc}
\epsfig{file=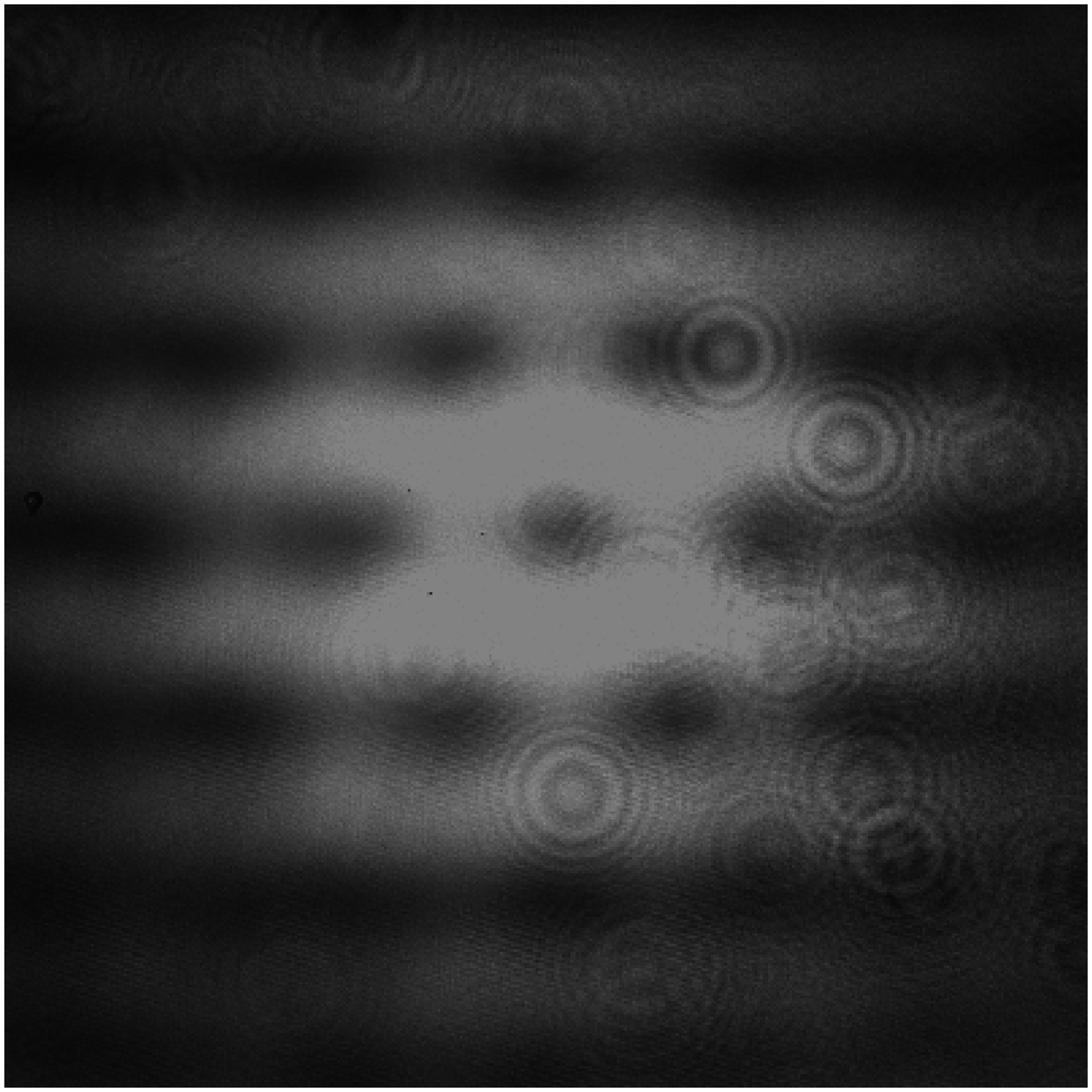,width=0.33\linewidth,clip=} & 
\epsfig{file=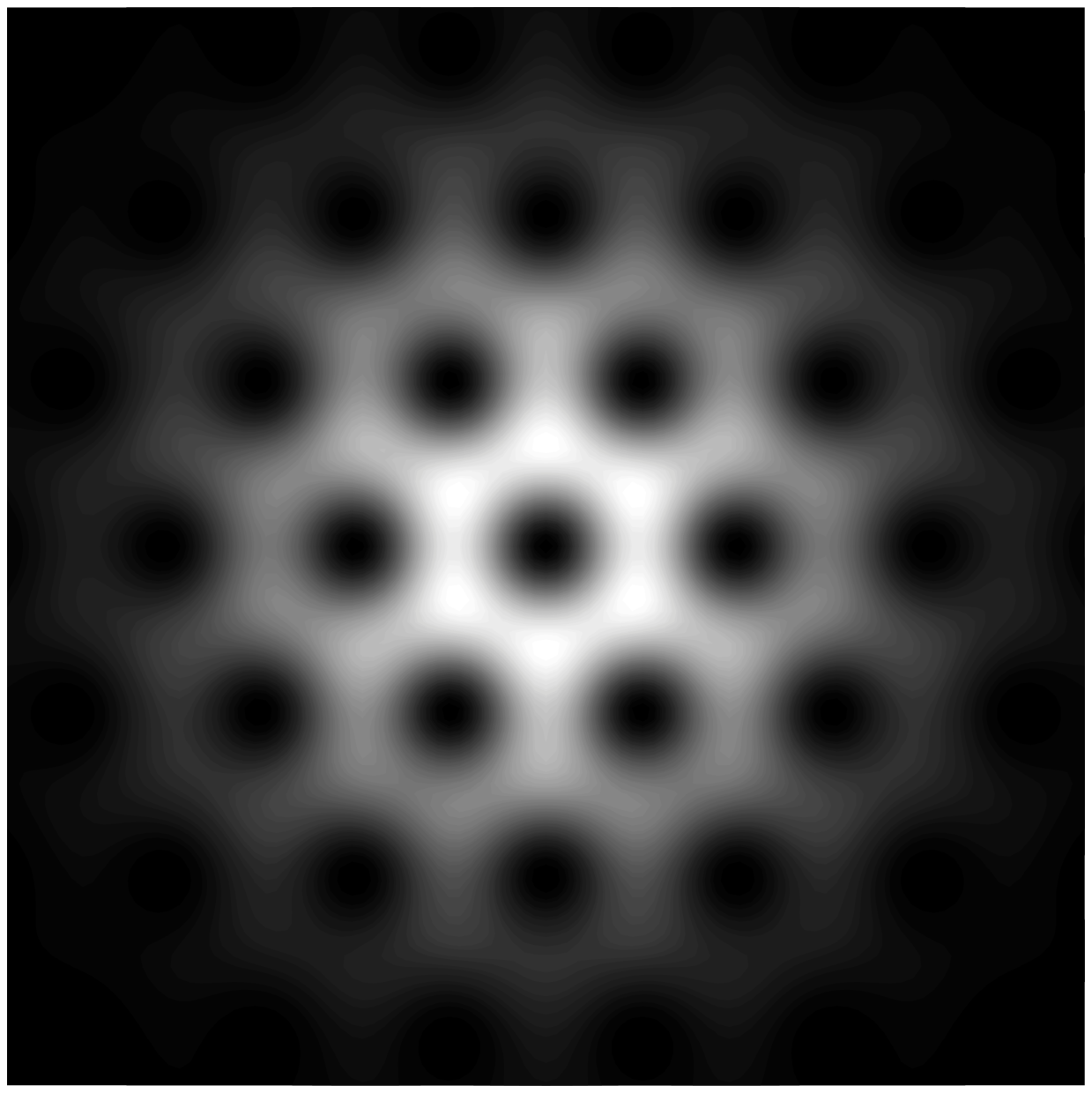,width=0.33\linewidth,clip=} \\
\epsfig{file=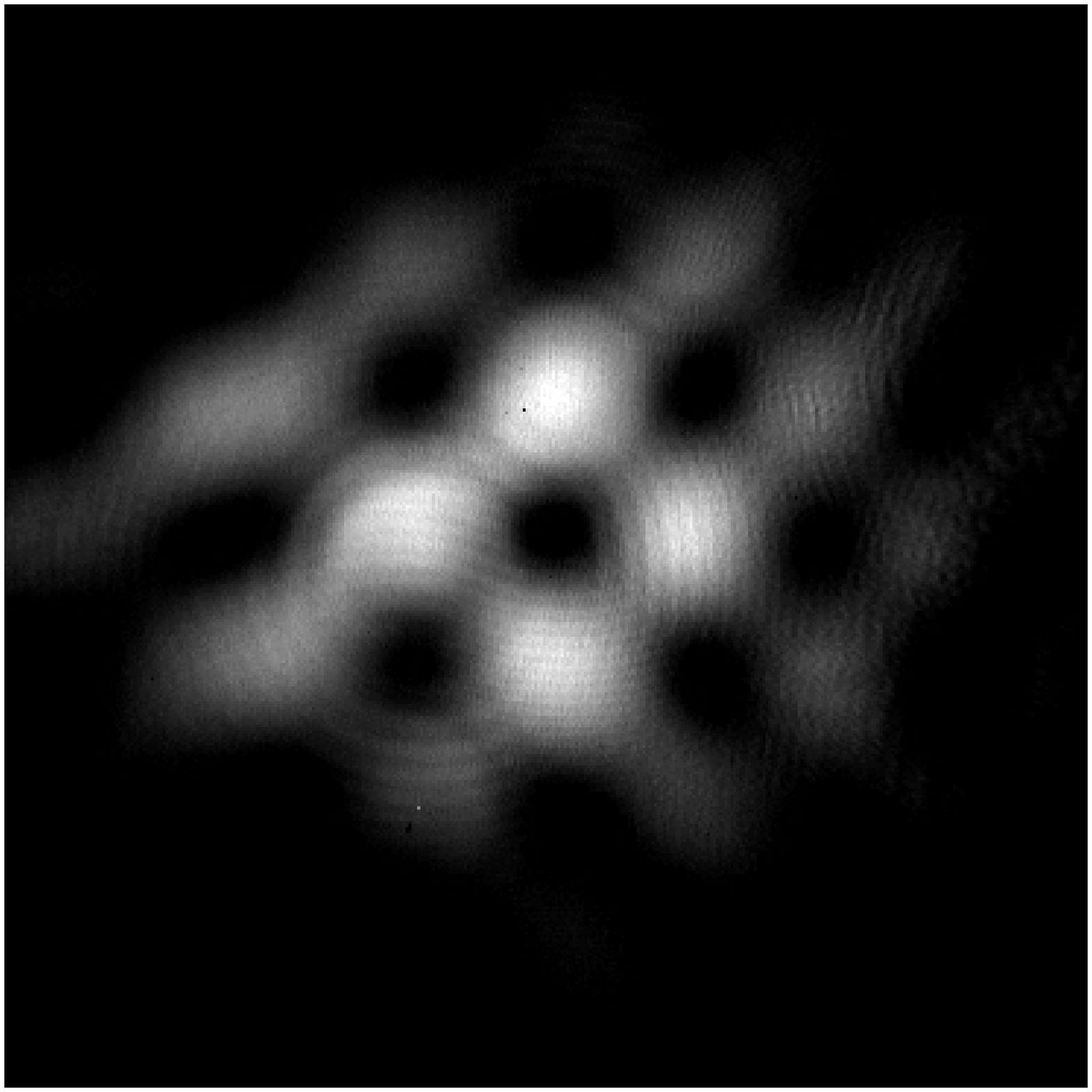,width=0.33\linewidth,clip=} &
\epsfig{file=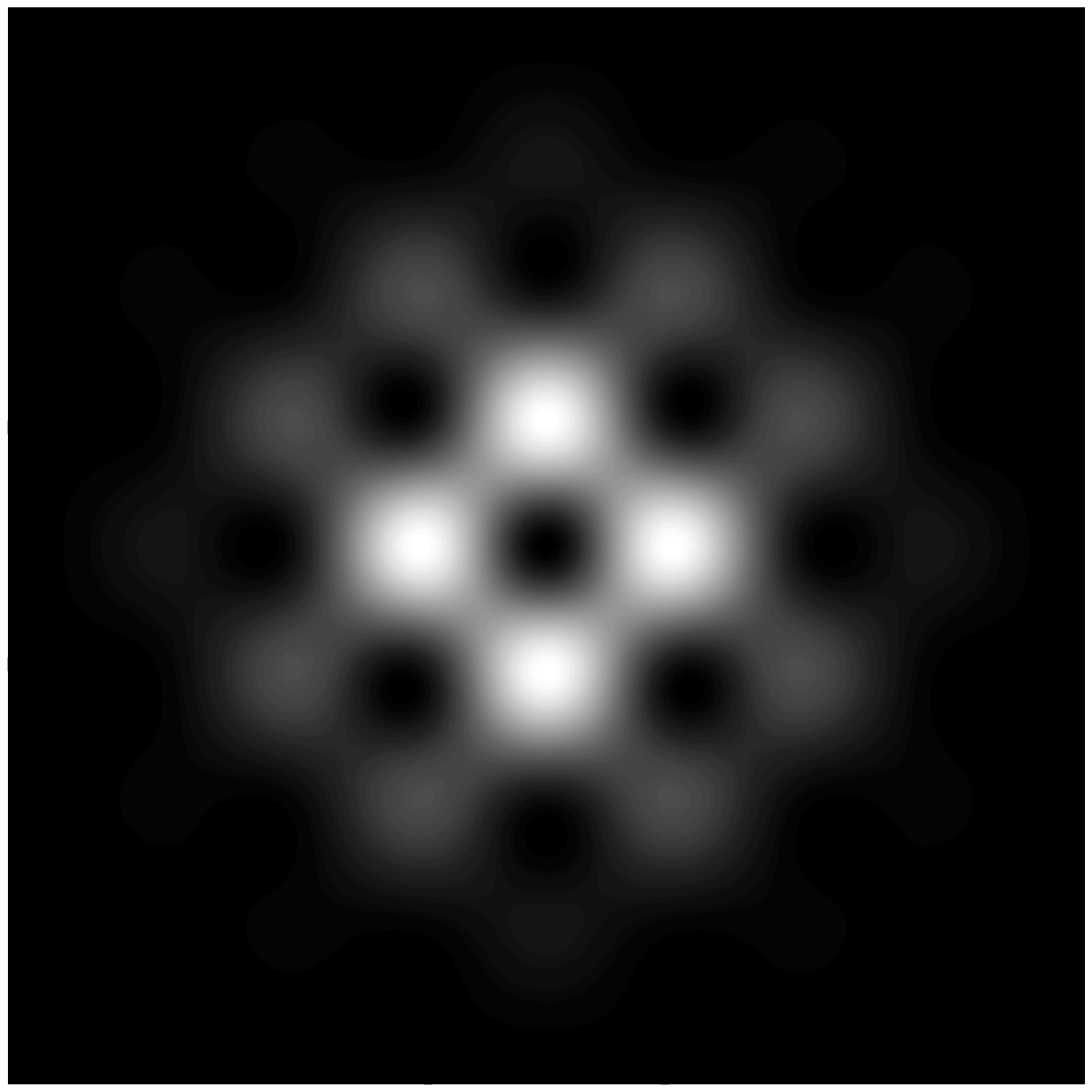,width=0.33\linewidth,clip=} \\
\epsfig{file=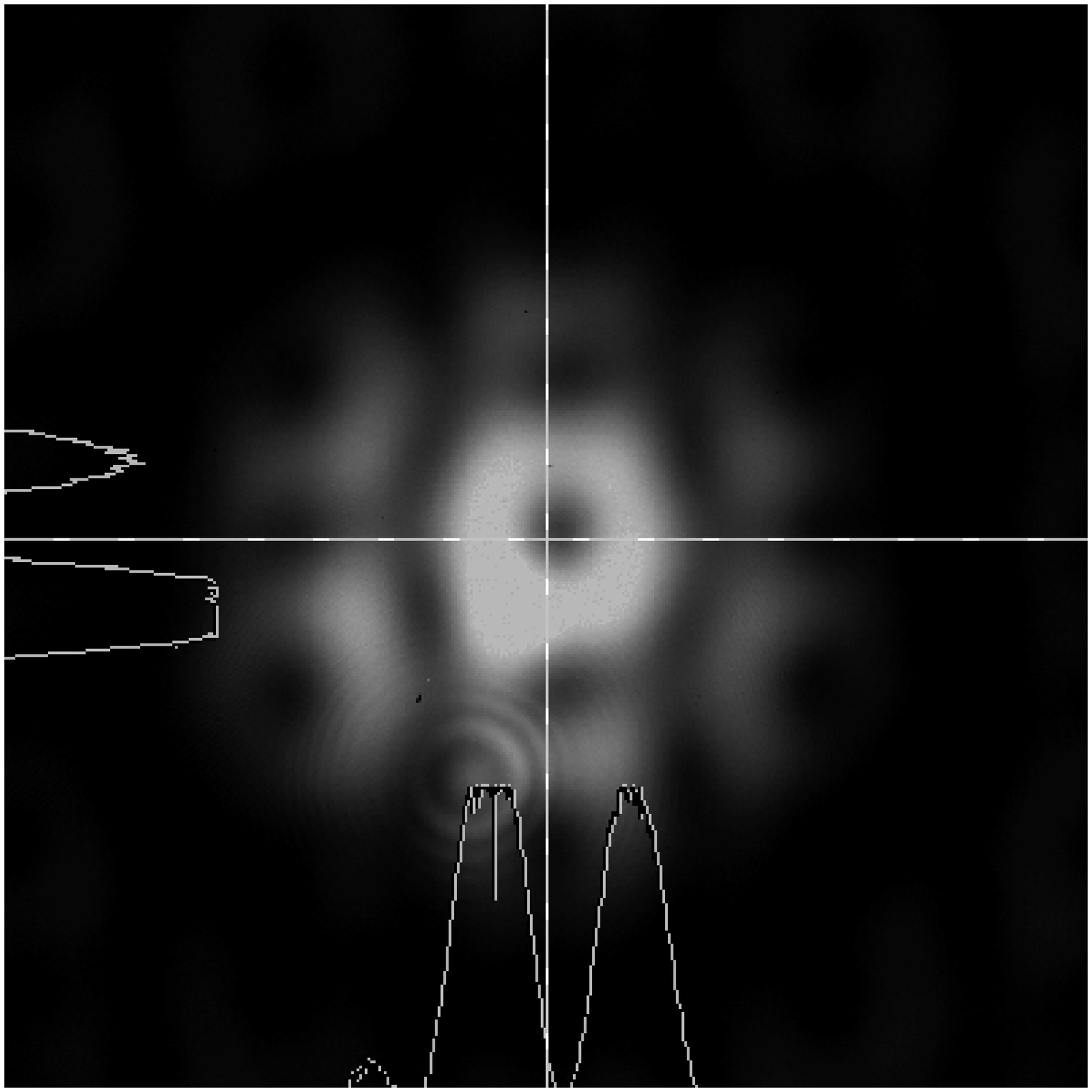,width=0.33\linewidth,clip=} &
\epsfig{file=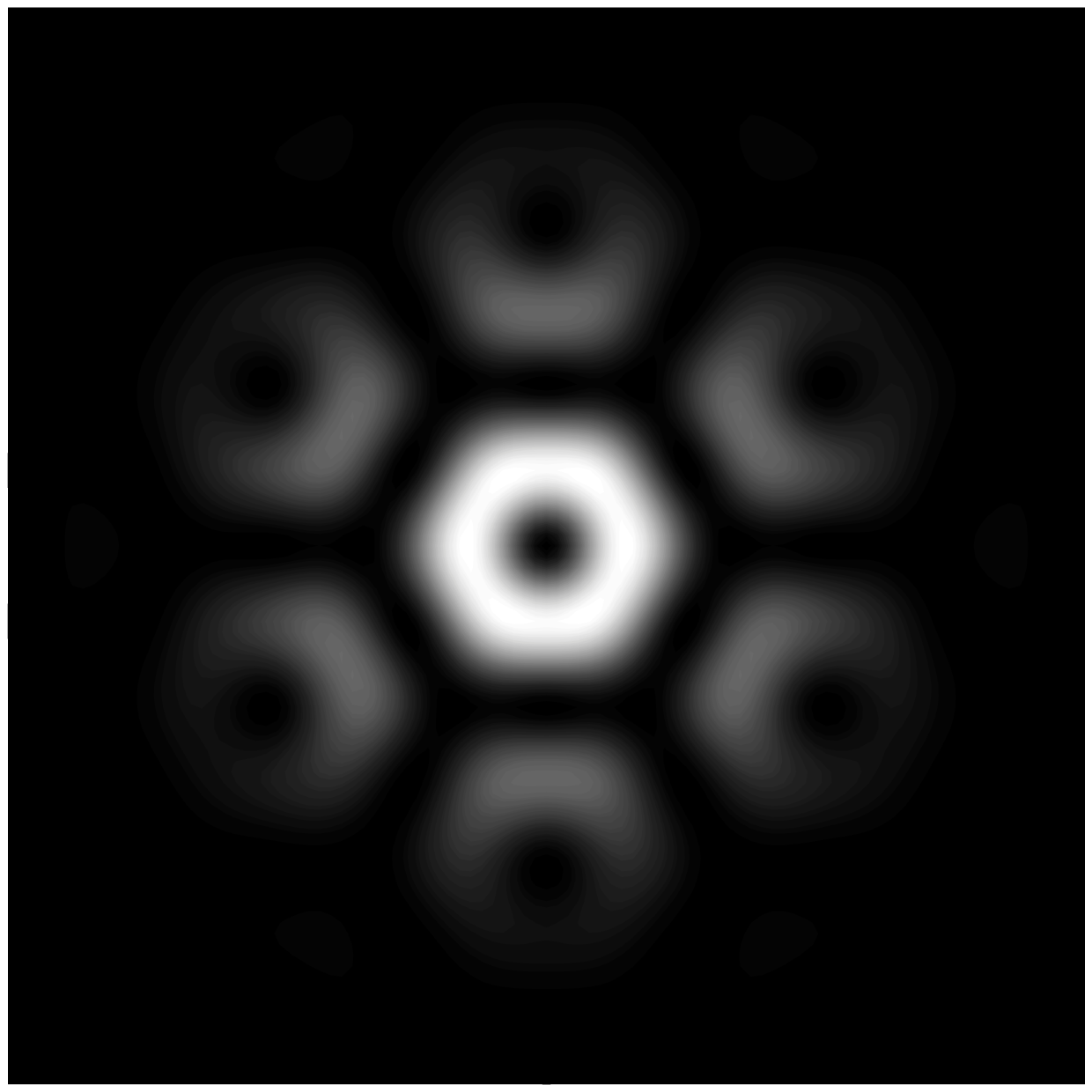,width=0.33\linewidth,clip=} 
\end{tabular}
\caption{Far field intensity profiles of effectively radially polarized beams.
The left panels correspond to the
experimental images and the right panels
represent the theoretical results. 
We consider a 3 (top row), 4 (middle row), and 6 (bottom row)
beam arrangement.
For $N=3$,
we take $w_0=32$ $\mu$m and $R=5 w_0$. 
For $N=4$, 
$w_0 = 42$ $\mu$m, and $R=2.7 w_0$.  For $N=6$, we have
$w_0= 43$ $\mu$m and $R=3.8 w_0$. Note that in general,
a radially polarized beam undergoes a discontinuity at
the origin, and thus the intensity must vanish there.
}
\label{all}
\end{figure}

The second method (for the case of $N=3$ and $6$), which is simpler for larger arrays, 
is shown in Fig.~\ref{figexp}.  
In this method the Nd:YAG is coupled into a polarization maintaining fiber.  
This signal is then split into multiple copies by means of a lithium
niobate waveguide 8-way splitter and 8-channel electro-optic modulator.  
Each path 
can then have a separate phase modulation to ensure proper phasing of the beams.  These signals are 
then propagated through fiber and coupled out and collimated by a 1 in. lens.  The lens is slightly 
overfilled so the Gaussian outputs of the fibers are truncated at $1.1 \times$  (the $e^{-1}$ radius).  


Experimental results for the  radial vector beam generation found good agreement with theoretical calculations, 
as illustrated in Fig.~\ref{all}. Clearly, the interference patterns and symmetry correlate well with the corresponding 
calculations in the three beam case. In particular, for $N=4$ (middle row), the central null surrounded by
a checkerboard peak structure. 
This pattern differs from the linear polarization by a rotation of $\pi/4$,
based on  a simple phase argument. 
The $N=6$ case is shown in the bottom row, where
clearly the intensity vanishes at the origin, followed by the formation of bright hexagonal rings. 
The three beam configuration is shown in the top row, where again there is satisfactory agreement between
the calculated and measured results. Any observed discrepancies may be reduced with
an appropriately incorporated feedback system. 
A substantial fraction of the energy in the plots of Fig.~\ref{all} is
contained within the first regions of high intensity peaks.
We illustrate this for the $N=4$ case where the net intensity within a circular region of radius $\rho$, $U(\rho)$, is calculated
to be approximately 
$U(\rho) \approx k_0 \rho z/(2 R) [k_0 \rho R/z - J_1(2 k_0 \rho R/z)]$. 
Inserting the appropriate parameters and normalizing to the intensity integrated over the entire image plane, yields
$50\%$ of the distribution is contained in the neighborhood of the first main peaks.

In conclusion,
it has been shown 
by employing two different configurations, 
and by carefully controlling the polarization, a central null can be created in the far field.  
In previous 
work, the central portion of a phased array of lasers typically has been a peaked function.  Now for 
the first time to our knowledge, a central null has been formed.  Similar results have been shown 
from single aperture lasers using an axicon to generate an annular mode inside a laser cavity.  
In the present case however, there is direct control over which mode will be generated from the same system.
As for scaling, there is no fundamental limitation on the number of beams used. 
Several methods are available to incorporate a greater number of beams, including
concentric beam placement (following the same beam-combining prescription described in this Letter) that could scale in roughly the same pattern
as a Bessel beam. It is also possible to use 
fiber lasers with active phasing, thus potentially scaling to many hundreds 
of beams\cite{shay}.
Our system also has a practical advantage over typical 
high power Gaussian 
beam applications which have the drawback of having the beam concentrated near the center, 
where the reflecting beam director (such as a telescope obscuration) resides.  Our device on the 
other hand, has no power propagating along the center axis of the beam,
and thus for high power applications requiring a center obscuration telescope beam director, our proposed system 
offers new advances.


We thank S. Feng for many useful discussions.

\end{document}